\documentclass[prl,twocolumn,showpacs,floatfix,amsmath,amssymb,superscriptaddress]{revtex4}
\usepackage[dvips]{graphicx}
\usepackage{latexsym}
\usepackage{graphicx}
\usepackage{times,psfrag,subfigure}
\usepackage{amsmath}
\usepackage{dcolumn}
\usepackage{latexsym,amsmath,amssymb,bm,euscript}
\def\etal{~\textit{et~al.}} 

\def\Hc{{\rm H.c.}}

\def\barpsi{\overline{\psi}}
\def\CsCuCl{$\text{Cs}_2\text{CuCl}_4$}

\begin{document}

\title{Algebraic vortex liquid in spin-1/2 triangular antiferromagnets:
Scenario for Cs$_2$CuCl$_4$}

\author{Jason Alicea}
\affiliation{Physics Department, University of California, 
Santa Barbara, CA 93106}
\author{Olexei I. Motrunich}
\affiliation{Kavli Institute for Theoretical Physics, 
University of California, Santa Barbara, CA 93106}
\author{Matthew P. A. Fisher}
\affiliation{Kavli Institute for Theoretical Physics, 
University of California, Santa Barbara, CA 93106}

\date{\today}

\begin{abstract}
Motivated by inelastic neutron scattering data on Cs$_2$CuCl$_4$, 
we explore spin-1/2 triangular lattice antiferromagnets 
with both spatial and easy-plane exchange anisotropies, the latter 
due to an observed Dzyaloshinskii-Moriya interaction.
Exploiting a duality mapping followed by a fermionization of the dual
vortex degrees of freedom, we find a novel ``critical" spin-liquid phase 
described in terms of Dirac fermions with an emergent global SU(4)  
symmetry minimally coupled to a non-compact U(1) gauge field.
This ``algebraic vortex liquid" supports gapless spin excitations and
universal power-law correlations in the dynamical spin 
structure factor which are consistent with those observed in Cs$_2$CuCl$_4$.
We suggest future neutron scattering experiments that should help
distinguish between the algebraic vortex liquid and other spin liquids 
and quantum critical points previously proposed in the context of 
Cs$_2$CuCl$_4$.
\end{abstract}

\maketitle

The search for two-dimensional (2D) spin liquids has been one of the most 
tantalizing pursuits in condensed matter physics.   
Among the most promising systems for realizing such states is the 
spin-1/2 triangular antiferromagnet, as both the low spin and 
geometric frustration suppress magnetic ordering.  This was
appreciated over three decades ago by Anderson, who postulated a
quantum-disordered ``resonating-valence-bond'' (RVB) ground state in the
spin-1/2 Heisenberg triangular antiferromagnet\cite{RVB}.
Anderson's RVB concept matured in the high-$T_c$ era, with
the triangular lattice often center stage.  Using slave bosons,
Sachdev explored an Sp($N$) generalization of the Heisenberg 
antiferromagnet and obtained a 
quantum-disordered ground state, the $Z_2$ spin liquid, which breaks no 
symmetries\cite{Sachdev}.
More recently, Moessner and Sondhi realized a $Z_2$ 
spin liquid in a quantum dimer model on the triangular 
lattice\cite{Moessner}.
By exploiting a correspondence between the triangular antiferromagnet 
and hard-core bosons in a magnetic field,
Kalmeyer and Laughlin introduced a ``chiral'' spin liquid\cite{KL}
which violates time-reversal symmetry.  Both the $Z_2$ and chiral
spin liquids admit gapped, fractionalized $s=1/2$
excitations---\emph{spinons}---which are bosonic in the former case and
``semionic'' in the latter.

\begin{figure}
\centerline{\includegraphics[width=6cm]{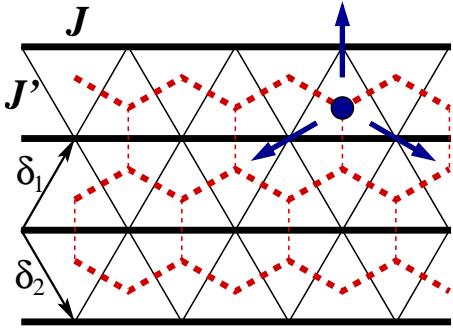}}
\caption{Triangular lattice and the dual honeycomb on which vortices reside.  
Spins shown illustrate a vortex.  The dominant
spin coupling/vortex hopping occurs along the corresponding bold links.  
With our convention the vortices are at half-filling, \emph{e.g.},
in the spiral state the vortex number is one (zero) on up (down) triangles.
} 
\label{lattice} 
\end{figure}

In spite of these theoretical advances, experimental spin-liquid 
candidates have only recently appeared.  One promising system is 
the spin-1/2 triangular antiferromagnet
\CsCuCl, which has anisotropic exchange energies $J = 4.3$K and $J' =
0.34 J$ (see Fig.\ \ref{lattice}).  Although this material 
develops long-range spiral order below
the Neel temperature $T_N = 0.62$K\cite{Coldea}, unusual 
features reminiscent of spin-liquid physics are manifested in its 
\emph{dynamics}, probed via inelastic neutron 
scattering\cite{Coldea,Coldea2}.  
Most notably, in addition to the sharp low-energy spin-wave peaks 
observed in the ordered phase, neutron scans at higher energies reveal
``critical'' power laws in the dynamical structure factor. 
This enhanced scattering persists in a range of temperatures 
above $T_N$, where the magnons are absent, and is suggestive of spinon
deconfinement characteristic of spin liquids.

This remarkable behavior has attracted much theoretical interest in 
\CsCuCl, and several scenarios for the origin of the anomalous
scattering have been proposed.  
Spin-wave theory\cite{Trumper, Veillette, Veillette2} 
and series expansion studies\cite{Zheng99, Zheng05}  
have yielded important quantitative 
connection with experiment.  Quasi-1D effects have been explored by
approaching the triangular lattice by coupling 1D 
chains\cite{Bocquet, Starykh}.  
Sachdev's slave boson approach was 
generalized to the anisotropic triangular antiferromagnet by 
Chung\etal\cite{Chung1}, and Isakov\etal\cite{Isakov}
explored the possibility that the \CsCuCl\ phenomenology may be 
controlled by a quantum critical point separating the $Z_2$
spin liquid and the spiral state.  Using slave fermions, 
Zhou and Wen\cite{ZhouWen} alternatively suggested the presence of a 
``critical'' algebraic spin liquid.

Here, we pursue a new theoretical approach to the triangular 
antiferromagnet.  We consider an easy-plane XXZ spin-1/2 system
reformulated in terms of \emph{fermionized vortex} degrees of freedom
using Chern-Simons flux attachment.  
Remarkably, this approach leads naturally to a new ``critical'' 
spin liquid---the ``algebraic vortex liquid''---which we explore and then 
apply to \CsCuCl.

\emph{Algebraic Vortex Liquid.} ---
Consider an easy-plane spin-1/2 antiferromagnet on the anisotropic 
triangular lattice shown in Fig.~\ref{lattice}.  
We return below to the appropriateness of the 
easy-plane assumption for \CsCuCl.
We follow closely a dual approach employing \emph{fermionized vortices}, 
developed for integer-spin systems in 
Ref.~\onlinecite{spin1}.
Implementing the standard duality mapping, one obtains a theory 
for interacting bosonic vortices on the dual honeycomb lattice 
(see Fig.~\ref{lattice}).  A crucial feature is that the vortices are at
\emph{half-filling} due to the spin frustration.
Moreover, because $S^z$ is a half-integer, the vortices ``see'' 
an average background of $\pi$ flux through each hexagon.
In terms of a vortex number operator $N_{\bf x}$ and its conjugate phase 
${\theta_{\bf x}}$, the vortex Hamiltonian is\cite{spin1}
\begin{eqnarray}
  {\mathcal H}_{\text{dual}} &=&
  - \sum_{\langle {\bf x x'} \rangle} t_{\bf x x'}  
  \cos(\theta_{\bf x} - \theta_{\bf x'} - a_{\bf x x'} - a^0_{\bf x x'})
  \nonumber \\
  &+& \sum_{\bf x x'} (N_{\bf x}-1/2) V_{\bf x x'} (N_{\bf x'}-1/2) 
   + {\mathcal H}_a ~.
  \label{Hdual} 
\end{eqnarray}
Here $V_{\bf x x'}$ encodes a logarithmic vortex repulsion; 
$a^0_{\bf x x'}$ is a static field satisfying 
$(\Delta \times a^0)_{\bf r} = \pi$, where 
$(\Delta\times a^0)_{\bf r}$ is a lattice curl around the hexagon 
encircling triangular lattice site ${\bf r}$; and
${\mathcal H}_a$ describes the dynamics of the dual gauge field
$a_{\bf x x'}$ residing on honeycomb links.
The $S^z$ spin component appears here as a dual gauge flux,
$S^z_{\bf r} \sim  (\Delta\times a)_{\bf r}/(2\pi)$.
The vortex hopping amplitudes $t_{\bf x x'}$ are anisotropic to reflect 
the lattice anisotropy of the exchanges.  Thus, vortices hop more easily
across weak spin links $J'$ than across strong links $J$.

Since the vortices are at half-filling, the dual theory appears
as intractable as the original spin model.  
One can, however, make significant progress by \emph{fermionizing} 
the vortices.  While similar approaches employing fermionized spins have
been pursued\cite{Lopez}, our treatment is appealing
because vortex density fluctuations are greatly suppressed by their
logarithmic repulsion.  
Following Ref.~\onlinecite{spin1}, we perform flux attachment 
to convert the bosonic vortices into fermions coupled to a 
Chern-Simons gauge field.
Taking a ``flux-smeared'' mean-field state, the Chern-Simons flux
averages to $2\pi$ per hexagon, which can be ignored on the lattice.  
Thus, the mean-field Hamiltonian describes half-filled fermionic
vortices hopping on the honeycomb in a background of 
$\pi$ flux (due to $a^0_{\bf x x'}$).  
The corresponding band structure reveals four gapless Dirac points.
Focusing on low-energy excitations in the vicinity of the four
nodes and including gauge fields, we obtain the Euclidean 
Lagrangian density
\begin{eqnarray} 
  {\mathcal L} &=& 
  \barpsi_\alpha (\partial \!\!\! / - i a \!\!\! / - i A \!\!\! / ) \psi_\alpha
  \nonumber \\
  &+& \frac{1}{2 e^2} (\epsilon_{\mu\nu\lambda}\partial_\nu a_\lambda)^2 
  + \frac{i}{4\pi} \epsilon_{\mu\nu\lambda} A_\mu \partial_\nu A_\lambda 
  + {\mathcal L}_{\rm 4f}.
  \label{L0} 
\end{eqnarray}
Equation~(\ref{L0}) describes four flavors of two-component Dirac 
fermions $\psi_{\alpha}$, $\alpha \!=\! 1,\dots,4$, minimally coupled to a 
non-compact U(1) gauge field $a_\mu$ and a Chern-Simons field $A_\mu$.
The gauge field $a_\mu$ mediates the logarithmic vortex repulsion, 
while the Chern-Simons terms implement the flux attachment.
Here and below, the nodal velocity anisotropies are absorbed in
the scaling of coordinates, while the Maxwell term is only schematic.
Symmetries of the original XXZ spin model preclude all possible 
mass terms in Eq.~(\ref{L0}).

The term ${\mathcal L}_{\rm 4f}$ represents symmetry-allowed 
four-fermion terms arising from short-range vortex interactions.
If sufficiently strong, such terms can drive spontaneous 
fermion mass generation, leading to various symmetry-broken 
states\cite{AVLlong}.  For example, giving the fermions a mass 
by introducing a staggered chemical potential on the honeycomb
induces a vortex density wave corresponding to the
spiral spin-ordered state.  On the other hand, the state with 
a spontaneously generated $\barpsi_\alpha \psi_\alpha $ mass 
({\it i.e.}, all fermion masses having the same sign) 
corresponds to the Kalmeyer-Laughlin chiral spin liquid.

In addition to broken-symmetry states, Eq.\ (\ref{L0}) 
allows us to access a new ``critical'' spin liquid.
To this end, we rewrite the Lagrangian in terms of
$\tilde a_\mu \equiv a_\mu + A_\mu$ and integrate out the 
Chern-Simons field to obtain ${\mathcal L} = {\mathcal L}_{\rm QED3}
 + O(\partial^3 \tilde{a}^2)$, with
\begin{equation}  
 {\mathcal L}_{\rm QED3} = 
    \barpsi_\alpha (\partial \!\!\! / - i \tilde{a} \!\!\! / ) \psi_\alpha
  + \frac{1}{2 e^2} (\epsilon_{\mu\nu\lambda} \partial_\nu 
                     \tilde{a}_\lambda)^2
  + {\cal L}_{4f}.
\label{QED3}
\end{equation}
Remarkably, up to higher-derivative terms which are 
henceforth dropped, we arrive at non-compact quantum 
electrodynamics in 2+1 dimensions (QED3), with $N = 4$ flavors of 
two-component Dirac fermions.\footnote{Dropping the higher-derivative 
term is justified by power counting, but is potentially dangerous since 
it changes the formal symmetries of the continuum action.  We assume 
that it is fine to do so when describing critical fluctuations.}
Physically, vortex density fluctuations are suppressed so strongly by
interactions that exchange statistics play only a minor role. 

QED3 has been widely studied ({\it e.g.}, see refs.\ in \cite{spin1}), 
and in the large-$N$ limit realizes a nontrivial stable critical 
phase.  For $N < N_c$, with some unknown $N_c$, it is believed that  
four-fermion terms become relevant and spontaneously generate fermion 
masses, destroying criticality (except at fine-tuned critical points).
Here, we proceed with the assumption that $N_c < 4$, implying the 
presence of a stable critical phase for our dual fermionized-vortex 
theory.  We now explore some of the properties of this 
``Algebraic Vortex Liquid" (AVL).

\begin{figure}
  \centerline{\includegraphics[width=5.5cm]{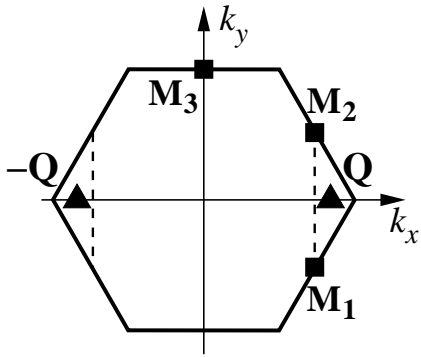}}
  \caption{Momenta in the triangular BZ at which dominant power-law
    correlations in the dynamic spin structure factor $S^{+-}$ occur in
    the AVL.  We predict the \emph{same power law} at all five
    momenta.  For $S^{zz}$, the dominant scattering is at
    $\pm{\bf Q}$ and has a faster-decaying power law.
    Dashed lines show $k_x=\pm\pi$ characteristic of a 1D system.}
  \label{MonopoleQs} 
\end{figure}

\emph{Critical spin correlations in the AVL.} ---
The AVL respects all symmetries of the microscopic spin system, 
exhibiting no magnetic or other types of order.
Since the Dirac fermions are gapless, power-law spin 
correlations are expected.  Consider first the in-plane spin components.
The spin raising operator $S^+_{\bf r}$ adds $S^z = 1$ and hence 
$2\pi$ dual gauge flux.  Near this flux insertion the fermionic 
Hamiltonian has four zero-energy modes, one for each flavor.  
Physical (gauge-invariant) states are obtained by occupying two of
these, so there are six distinct such ``monopole insertions.''
Following the procedure of Ref.~\onlinecite{spin1}, we determine the
momenta carried by the monopoles: two occur at the spiral
ordering wave vectors $\pm{\bf Q}$ and three occur at the midpoints
${\bf M_{1,2,3}}$ of the Brillouin zone (BZ) edges (see Fig.\
\ref{MonopoleQs}).  
Numerical diagonalization suggests that the sixth monopole, 
which carries zero momentum, does not have the same symmetry as 
$S^+$\cite{AVLlong}, and
thus by itself will not contribute to the in-plane spin correlations.  

Monopole insertions are known to have nontrivial power-law correlations 
in the large-$N$ QED3 theory.  The scaling dimension, $\Delta_{\rm m}$,
of the monopoles can be estimated from the leading large-$N$ 
result\cite{BKW}, $2 \Delta_{\rm m} \approx 0.53 N$.  
A naive extrapolation to $N=4$ yields 
$\eta_{\rm m} \equiv 2\Delta_{\rm m} - 1 \approx 1.12$.
Since all monopoles have the same scaling dimension, $S^+$ is 
expected to exhibit the same universal power-law correlations at all 
five momenta ${\bf K}_j$ shown in Fig.\ \ref{MonopoleQs}.  
Specifically, near each ${\bf K}_j$, the dynamic spin structure factor 
is predicted to scale as
\begin{eqnarray}
{\cal S}^{+-} ({\bf k} \!=\! {\bf K}_j + {\bf q},\, \omega) 
= A_{{\bf K}_j} \; 
\frac{\Theta(\omega^2 - {\bf q}^2)}
     {(\omega^2 - {\bf q}^2)^{1-\eta_{\rm m}/2}} ~.
\label{Scrit}
\end{eqnarray}
The amplitudes $A_{{\bf K}_j}$ are sensitive to short-distance physics 
and can differ significantly among the five wave vectors, particularly 
in the anisotropic system.
With $J \gg J^\prime$ the amplitude at wave vector 
${\bf M_3}$ with $k_x=0$ is expected to be much suppressed compared to 
the other four momenta, as the latter are near $k_x=\pi$ where the
dominant antiferromagnetic correlations occur along the nearly decoupled
chains.

The $S^z$ spin correlation behaves rather differently.
At zero momentum, the correlation is that of the conserved dual gauge 
flux.  However, a more prominent power law occurs at the spiral 
ordering wave vectors $\pm{\bf Q}$.  
These arise because an expression for $S^z$ in terms of continuum fields 
allows a term 
$e^{i \bf Q \cdot r} \barpsi_\alpha {\rm W}_{\alpha \beta} \psi_\beta + \Hc$
with a fermionic bilinear $\barpsi \hat{\rm W} \psi$ whose correlation
is enhanced by gauge field fluctuations.  
Here, $\hat{\rm W}$ can be obtained by considering a perturbation to the 
fermionic hopping Hamiltonian that adds static gauge flux through the hexagons 
modulated at wave vector ${\bf Q}$.
The ${\cal S}^{zz}({\bf k}, \omega)$ dynamical spin structure factor
near $\pm{\bf Q}$ has a form similar to Eq.~(\ref{Scrit}), 
except the corresponding anomalous dimension is that of an enhanced 
fermionic bilinear in QED3.
Using the $1/N$ result\cite{Rantner}, 
$\eta_{\rm enh} \approx 3 - 128/(3\pi^2 N)$,
we estimate $\eta_{\rm enh} \approx 1.92$.
At other wave vectors the dynamical correlation 
${\cal S}^{zz}({\bf k}, \omega)$ exhibits subdominant power laws.

\emph{ \CsCuCl\ Hamiltonian.} ---
To address the possible experimental relevance of the AVL phase, 
we now consider the measured \CsCuCl\ Hamiltonian.
Besides the Heisenberg exchange with $J = 4.3$K, $J' = 0.34 J$,
there is also a Dzyaloshinskii-Moriya (DM) interaction
\begin{equation}
  H_{\rm DM} = -\sum_{\bf r} {\bf D}\cdot {\bf S}_{\bf r}\times
  ({\bf S}_{{\bf r} + {\bm \delta}_1} 
   + {\bf S}_{{\bf r} + {\bm \delta}_2}).
  \label{HDMspin}
\end{equation}
The vectors ${\bm \delta}_{1,2}$ connect spins on neighboring 
chains as in Fig.~\ref{lattice}, and ${\bf D} = D{\bf \hat z}$ 
is oriented perpendicular to the triangular layers, with $D = 0.053 J$.
Although there is a small interlayer coupling\cite{Veillette}, we focus on the 
2D system.

Significantly, the DM term provides an easy-plane anisotropy, breaking 
the SU(2) spin symmetry of the Heisenberg exchange down to U(1),
and also violates inversion symmetry ${\bf r}\to {-\bf r}$.
Thus for $T<T_N = 0.62$K the DM term determines both the ordering plane, 
which coincides with the triangular layers, and the sign of the 
ordering wave vector, which along with $D$ changes sign from one 
layer to the next.
Despite the small value of $D$, the easy-plane anisotropy is amplified 
since the DM interaction is not frustrated near the dominant 
antiferromagnetic wave vector along the chains (whereas the $J'$
coupling is frustrated).
To quantify this, we briefly consider a 
{\it classical} 2D spin system with \CsCuCl\ parameters.  
Without the DM term, a $J-J'$ Heisenberg spin
system remains disordered at all temperatures.  With DM 
coupling the system has only U(1) spin symmetry and 
thus exhibits a low-temperature phase with quasi-long-range order (QLRO).  
We performed a classical Monte Carlo study and found this 
transition at $T_c \approx 0.27\, J S^2$.  A simple classical
ground state analysis indicates that about half of the 
``phase stiffness'' originates from the DM coupling itself. 
Taking $S^2 = 3/4$ appropriate for spin-1/2, we estimate that a 
single layer would obtain QLRO below $T_c = 0.84$K.  
Once each layer has QLRO, an arbitrarily small interlayer coupling 
would induce 3D long-range order, even though the interlayer coupling 
is frustrated in \CsCuCl\ due to the alternating sign of 
$D$\cite{Veillette}.  This suggests that the observed spiral 
ordering at $T_N$ in \CsCuCl\ is primarily driven by the 
easy-plane character of spins in each layer.  As the classical 
treatment neglects quantum fluctuations, it is reasonable 
that the estimated $T_c$ somewhat exceeds $T_N$. 
These considerations suggest that vortices, required to drive the 2D 
ordering, acquire integrity as degrees of freedom in \CsCuCl.

\begin{figure}
\centerline{\includegraphics[width=5.5cm]{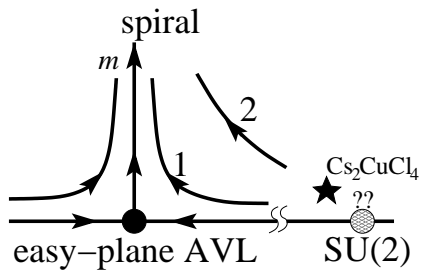}}
\caption{Schematic renormalization group flows near the AVL fixed point in the
presence of the DM interaction.  The vertical axis is the
inversion-breaking fermion mass $m$ arising from the 
DM term.  The bare \CsCuCl\ Hamiltonian is indicated with the star.
For flow trajectory 1, the dynamics at intermediate energy scales
is governed by the AVL theory.  For trajectory 2, intermediate
energy scales are controlled by an (unknown) SU(2)-invariant 
criticality.}
\label{scenario}
\end{figure}

\emph{Scenarios for AVL in \CsCuCl.} ---
To specifically address the applicability of the AVL to \CsCuCl,
we consider the effect of the DM term on the AVL.
In the fermionized-vortex Lagrangian~(\ref{L0}), the DM interaction
appears as an inversion-breaking fermion mass term corresponding to a
staggered vortex chemical potential, with bare mass $m \sim D$.  
This mass drives the 
system to the spiral state as indicated by the vertical flow in 
Fig.~\ref{scenario}, with chirality dictated by the sign of $D$.  
Thus, with the DM interaction the observed spiral 
ground state of \CsCuCl\ emerges 
naturally out of the critical AVL.  

Since the DM term induces an easy-plane spin anisotropy 
\emph{and} breaks inversion symmetry, applicability of the AVL 
to \CsCuCl\ 
requires a delicate balance.  For the AVL to apply on intermediate 
energy scales, the DM interaction must first produce sufficient 
easy-plane anisotropy for the description in terms of vortices to 
be appropriate, before destabilizing the AVL state toward the 
spiral order.
This is scenario~1 in the schematic flow diagram of Fig.~\ref{scenario}.
The alternative scenario~2 does not approach the easy-plane fixed
point but is driven directly to the magnetic order.
In the latter case, intermediate energy scales may be governed
by an (unknown) SU(2)-invariant criticality indicated with question 
marks in the figure.
Below, we pursue the consequences of scenario~1.

\emph{Comparison with experiments.} ---
The spin dynamics in \CsCuCl\ was measured in neutron scattering
experiments by Coldea\etal\cite{Coldea,Coldea2}.
For $T < T_N$, \CsCuCl\ has well-defined spin waves, gapless near
zero momentum and observed with a 
small gap near the ordering wave vectors $\pm{\bf Q}$, presumably due to 
weak violations of SU(2) spin symmetry by the DM term. 
Experiments also see fairly small spin-wave gaps at momenta
${\bf M_{1,2}}$ of Fig.~\ref{MonopoleQs}.
(Series expansion studies of the Heisenberg model show that the gaps 
near these momenta and also ${\bf M_{3}}$ deviate strongly from 
spin-wave theory\cite{Zheng05}.)
Broad continuum scattering is observed above the spin-wave 
gaps near momenta $\pm{\bf Q}$ and ${\bf M_{1,2}}$, and persists even 
for $T>T_N$.  Notably, Ref.~\onlinecite{Coldea2} reports power-law 
line shapes in scans near these momenta 
(scans J and G in Ref.~\onlinecite{Coldea2}), each with the {\it same} 
exponent, $\eta_{\rm exp} = 0.74$.
The AVL also admits gapless spin-1 excitations with power-law 
scaling at these momenta, with the leading 
estimate of $\eta_{\rm m}=1.12$ somewhat larger than experiment.
These momentum space locations of enhanced scattering are determined by 
physics on the scale of several lattice spacings, and provide important 
constraints on theory.
The AVL theory captures this aspect of the \CsCuCl\ phenomenology
rather well.

Since the above data are near $k_x = \pi$, which is the dominant 
wave vector in the quasi-1D limit, some caution is necessary.  
Specifically, with $J \approx 3 J^\prime$ strong contributions on
intermediate energy scales from antiferromagnetic correlations 
along the chains are not unexpected.  A measurement midway between 
points ${\bf Q}$ and ${\bf M_1}$ would help determine the
transverse $k_y$-dependence of the continuum scattering 
and clarify whether it is meaningful to speak of enhanced scattering 
near discrete momenta in the 2D BZ.
The AVL also predicts enhanced ${\cal S}^{+-}({\bf k}, \omega)$ 
near ${\bf M_3}$, albeit with a smaller amplitude.  
Being less influenced by quasi-1D effects, 
further measurements near ${\bf M_3}$ would be useful.
Polarized neutron experiments to search for the distinct
easy-plane character of the AVL would also be very interesting.  

Competing 2D theoretical proposals with full SU(2) spin symmetry include
an algebraic spin liquid (``U1C'') proposed by 
Zhou and Wen\cite{ZhouWen} and the quantum critical point (QCP) 
scenario of Isakov\etal\cite{Isakov}.  
Each makes distinct predictions for the momenta of the 
low-energy spin-1 excitations, which also differ from the AVL.
Such characterizations can in principle be used to discriminate
among different theories, although the limited energy window 
for observing the continua and the material's strong anisotropy 
are unavoidable complications.
Further experimental and theoretical studies should help clarify the 
true ``spin-liquid" nature of this interesting material.

\emph{Acknowledgments.} --- 
We would like to thank Leon Balents, T. Senthil and Martin Veillette 
for sharing their insights, and especially Michael Hermele for an 
initial collaboration.
This work was supported by the National Science Foundation (J.\ A.)
through grants PHY-9907949 (O.~I.~M.\ and M.\ P.\ A.\ F.) and 
DMR-0210790 (M.\ P.\ A.\ F.).


\end{document}